\documentclass[a4paper]{article}

\usepackage[english]{babel}
\usepackage[utf8x]{inputenc}
\usepackage[T1]{fontenc}
\usepackage{float}
\usepackage[a4paper,top=3cm,bottom=2cm,left=3cm,right=3cm,marginparwidth=1.75cm]{geometry}
\usepackage{centernot}
\usepackage{amsmath}
\usepackage{graphicx}
\usepackage{url}
\usepackage[colorinlistoftodos]{todonotes}
\usepackage[colorlinks=true, allcolors=blue]{hyperref}
\usepackage{authblk}
\usepackage[noadjust]{cite}
   
\usepackage{url}

\usepackage{graphicx, caption, subcaption}
\title{\textbf{Enhanced Photon Squeezing in Two-Photon Dicke Model}}

\author{Priyankar Banerjee\textsuperscript{1}\thanks{Email: p.banerjee@iitg.ac.in},\quad Deepti Sharma\textsuperscript{2}\thanks{Email: deepti.kapil.sharma@gmail.com}\quad and Aranya B. Bhattacherjee\textsuperscript{3, 4}\thanks{Email: aranyabhuti@hyderabad.bits-pilani.ac.in}}
\affil{\small{\textsuperscript{1}Indian Institute of Technology, Guwahati, Assam 781039, India\\\textsuperscript{2}School of Physical Sciences, Jawaharlal Nehru University, New Delhi 110067, India\\\textsuperscript{3}Birla Institute of Technology and Science Pilani, Hyderabad Campus, Telangana 500078, India\\\textsuperscript{4}University of Science and Technology,  Meghalaya 793101, India}}
\begin{document}
\maketitle
\begin{abstract}
We explore the phenomena of quadrature squeezing of photons in the Two-Photon Dicke Model under the mean-field approximation in the thermodynamic limit. The strength of photon squeezing is maximized in the region where the coupling strength is of the same order of magnitude as one of the fundamental frequency of the system. This particular region is termed as the unbounded region. The squeezing of the photonic quadratures can be observed only in the superradiant phase where the squeezing is well beyond the standard quantum limit both near the quantum critical point as well as the unbounded region. However, prolonged squeezing is only obtained in the latter case. Furthermore, we explore the critical behavior of photon squeezing near the unbounded region.
\end{abstract}
\section{Introduction} \label{sec:intro}
Squeezed states with reduced fluctuations preserving Heisenberg's uncertainty relation find applications in demanding areas such as high-precision measurements \cite{PhysRevD.23.1693,SCHNABEL20171}, quantum information processing \cite{yonezawa2008continuousvariable,Furusawa:09} and quantum communication \cite{PhysRevA.93.032308} which has attracted wide-spread attention in recent times \cite{Vahlbruch_2007,PhysRevLett.95.211102,PhysRevLett.88.231102,doi:10.1126/science.1086489}. Squeezed spins in atomic systems \cite{PhysRevA.68.013807, PhysRevA.76.043621, PhysRevResearch.2.033504} and squeezed bosons with squeezed states of light are useful in reducing quantum noise \cite{doi:10.1119/1.14327,Dwyer2013QuantumNR,Assaf:02,6988398, 1987} which arise during measurement processes \cite{PhysRevA.36.1249}. Apart from achieving squeezing, improving its strength and maintaining squeezing for a longer period of time, is among the many other aspects of studying squeezing \cite{PhysRevA.102.033702}. 

The phenomena of squeezing has been widely studied in different quantum optical models, which include one- and two-axis spin squeezed states under different approximations \cite{Jin_2009,PhysRevResearch.2.033504,2017,doi:10.1142/S021797921750062X}. Multi-qubit models like the Dicke Model \cite{PhysRev.93.99,BRANDES2005315} (which exhibits a quantum phase transition from the normal to the superradiant phase \cite{PhysRevA.92.033817,PhysRevLett.90.044101,PhysRevE.67.066203,PhysRev.188.1976,HEPP1973360}) has also been studied for the dynamics of squeezing across the phase boundary. The Two-Photon Dicke Model  \cite{GERRY1989263,Klimov_1999} involves N-qubits interacting with the cavity mode via a second-order process. Such second-order processes are described in various systems as quantum dots \cite{PhysRevB.73.125304,PhysRevB.81.035302} and Rydberg atoms \cite{PhysRevLett.88.143601,PhysRevLett.110.090402} by two-photon interactions. As shown by Garbe et al. \cite{PhysRevA.95.053854} under mean-field approximation, photon squeezing is obtained in the superradiant phase of the Two-Photon Dicke Model in the thermodynamic limit. 

In this paper, we aim to study and analyze the pattern of quadrature squeezing of the photons in the superradiant phase of the Two-Photon Dicke model under mean-field and Holstein Primakoff approximation. Unlike in the previous calculation \cite{PhysRevA.102.033702}, where squeezing is obtained close to the critical point in either of the two phases, here in contrast, squeezing is obtained only in the superradiant phase \cite{PhysRevA.95.053854}. This finding tells us that for this system, even though some degree of squeezing is present near the quantum critical point, its magnitude is maximum and prolonged only for a particular value of the coupling parameter (i.e. near the unbounded region). Here, we find maximized squeezing away from this critical point, which is an interesting observation. Also it might not always be feasible to keep the system near the critical phase boundary experimentally. Hence, these results are of  much practical importance.\\
This paper is divided into three segments. In Sec. \ref{sec:Model} we discuss the Two-Photon Dicke Model, revisit its mean-field analysis in Sec. \ref{sec:MFT} and then proceed to diagonalize the mean-field hamiltonian. In Sec. \ref{sec:squeezing}, we look at the time evolution of the squeezing parameter and find its behavior very close to the unbounded region. Finally, we conclude in Sec.  \ref{sec:conclusion} with a summary.
\section{Model and Hamiltonian  \label{sec:Model}}
We consider a system where a single cavity mode interacts with N identical two-level atoms via two-photon interactions. The hamiltonian to describe such a system is given by that of the Two-Photon Dicke Model: \\
\begin{equation}\label{Ham}
\begin{split} 
\hat{H}&=\omega\hat{a}^{\dagger}\hat{a} + \epsilon \hat{S}_z + \frac{\tilde{g}}{N}\hat{S}_x(\hat{a}^{{\dagger}2}+\hat{a}^2)\\
\end{split}
\end{equation}
where, $\omega$ is the frequency of the cavity mode, $\epsilon$ is the transition frequency of the two-level atoms in the cavity and $\tilde{g}$ is the coupling parameter. The spin operators have been defined as $\hat{S}_m = \frac{1}{2}\sum_{k=1}^{N}\hat{\sigma}_k^{m}$, where $m = {x,y,z}$. 
\subsection{Mean-Field Approximation} \label{sec:MFT} 
First, we will review briefly the mean-field results of Ref. \cite{PhysRevA.95.053854} in the large-N limit, then we will look at the squeezing dynamics we have derived. The Two-Photon Dicke Model undergoes quantum phase transition from normal to superradiant phase, provided $\tilde{g}<\omega/2$. Following Holstein-Primakoff Transformation \cite{PhysRev.58.1098}, the spin operators can be approximated as:
$\hat S_x = \frac{\sqrt{N-\hat{b}^{\dagger}\hat{b}}}{2}(\hat{b}^{\dagger}+\hat{b})$ and $\hat{S}_z = \hat{b}^{\dagger}\hat{b} - \frac{N}{2}$. \\
Using these transformed operators, the hamiltonian can now be written as:
\begin{equation}\label{Ham2}
    \begin{split}
        \hat{H}=\omega\hat{a}^{\dagger}\hat{a} + \epsilon \Big(\hat{b}^{\dagger}\hat{b} - \frac{N}{2}\Big) +\frac{g\sqrt{N-\hat{b}^{\dagger}\hat{b}}}{N}(\hat{b}^{\dagger}+\hat{b})(\hat{a}^{{\dagger}2}+\hat{a}^2)
    \end{split}
\end{equation}
Where, $g =\tilde{g}/2$. Defining the order parameter of the system as: $\beta = \left \langle b\right \rangle$, we can see that there is a quantum phase transition on variation of $\beta$. Now taking  $\hat{b} = \beta + \delta \hat{b}$ and $\hat{b}^{\dagger} =\beta^{\star} +\delta \hat{b}^{\dagger}$ and hence neglecting the spin fluctuations ($ \delta\hat{b}$ and $\delta\hat{b}^{\dagger}$), we see $\hat{b}^{\dagger}\hat{b} =(\beta^{\star} +\delta\hat{b}^{\dagger}) ( \beta + \delta \hat{b})=|\beta|^2$ and $\hat{b}^{\dagger}+\hat{b} =\beta^{\star} + \beta$. \\
Hence, the hamiltonian \eqref{Ham} becomes
\begin{equation}
\begin{split}
\hat{H}&=\omega\hat{a}^{\dagger}\hat{a} + \epsilon\Big(|\beta|^2 - \frac{N}{2}\Big) +g_{\beta} (\hat{a}^{{\dagger}2}+\hat{a}^2)\\
\end{split}
\end{equation}
where, $g_{\beta} = \frac{g\sqrt{N-|\beta|^2}}{N}(\beta^{\star} + \beta)$. \\

Now, this hamiltonian is of general form:
\[\hat H  = c_1  a^\dagger a  + c_2[a^\dagger a^\dagger + aa] +c_3 \]where, $
c_1 = \omega $, $c_2 = g_{\beta}$, $c_3 = \epsilon(|\beta|^2-\frac{N}{2})$.
\subsection{Diagonalization}
Bogoliubov Transformation of a form:  $\hat U_a = e^{-\frac{\theta_a}{2} (\hat a^{\dagger}\hat a^{\dagger}-\hat a\hat a)}$ is used to diagonalize a hamiltonian of this form \cite{PhysRevA.102.033702}. \\
The diagonalized hamiltonian is:
\begin{equation}\label{diag1}
\begin{split}
\hat{U}^{\dagger}_a\hat{H}\hat{U}_a & =  (\hat a\hat a^\dagger + \hat a^\dagger \hat a) (-\frac{c_1}{2}\cosh 2 \theta_a + c_2 \sinh 2\theta_a)+ c_3- \frac{c_1}{2}\\
\Rightarrow \mathcal{H}&= \omega_a \Big( \hat{a}^{\dagger}\hat{a}+ \frac{1}{2}\Big) + \epsilon\Big(|\beta|^2 - \frac{N}{2}\Big) - \frac{\omega}{2}
\end{split}
\end{equation}
Putting back the values of $c_3$ and $c_1$ in \eqref{diag1}, we see that $\omega_a = \omega\cosh 2\theta_a - 2g_{\beta} \sinh 2\theta_a$, where  $\theta_a = \frac{1}{2}\tanh^{-1}(\frac{2g_{\beta}}{\omega})$. \\
The ground state energy of the system is:
\begin{equation}
\begin{split}
E_g &= \left \langle 0 |\mathcal{H} | 0 \right \rangle\\
&=\frac{\omega_a}{2}+ \epsilon\Big(|\beta|^2 - \frac{N}{2}\Big) - \frac{\omega}{2}\\
& = \frac{\cosh 2 \theta_a}{2 \omega} (\omega^2-4g_{\beta}^2)+\epsilon \Big(|\beta|^2 - \frac{N}{2}\Big)  - \frac{\omega}{2}\\
\end{split}
\end{equation}
Now, the value(s) of $\beta$ for which the value of $E_g$ is minimum is calculated by setting $\frac{\partial E_g}{\partial \beta}|_{\beta = \beta_0} = 0$. 
\begin{equation}
\begin{split}
\beta & = 0 \text{ for  } g \leq g_t=\sqrt{\frac{\omega \epsilon N}{4}}\\
& =\pm \sqrt{\frac{N}{2}\Big(1-\sqrt{\frac{1-\mu}{4\mu^2\lambda^2 - \mu}}\Big)} \text{ for  } g > g_t=\sqrt{\frac{\omega \epsilon N}{4}}\\
\end{split}
\end{equation}
Where, $\lambda = \frac{\omega}{2 \epsilon  N}$ and $\mu = \frac{4g^2}{\omega^2}$. Here, the quantum critical point is given as $g_t=\sqrt{\frac{\omega \epsilon N}{4}}$.
\begin{equation}\label{qpt}
\begin{split}
&\text{For}\quad g \leq g_t=\sqrt{\frac{\omega \epsilon N}{4}} \text{,}\quad \text{Normal Phase}\\
&\text{For}\quad g>g_t=\sqrt{\frac{\omega \epsilon N}{4}} \text{,}\quad \text{Superradiant Phase}\\
\end{split}
\end{equation}
\begin{figure}[H]
\centering
\includegraphics[width= 12.5cm,height=\textheight,keepaspectratio]{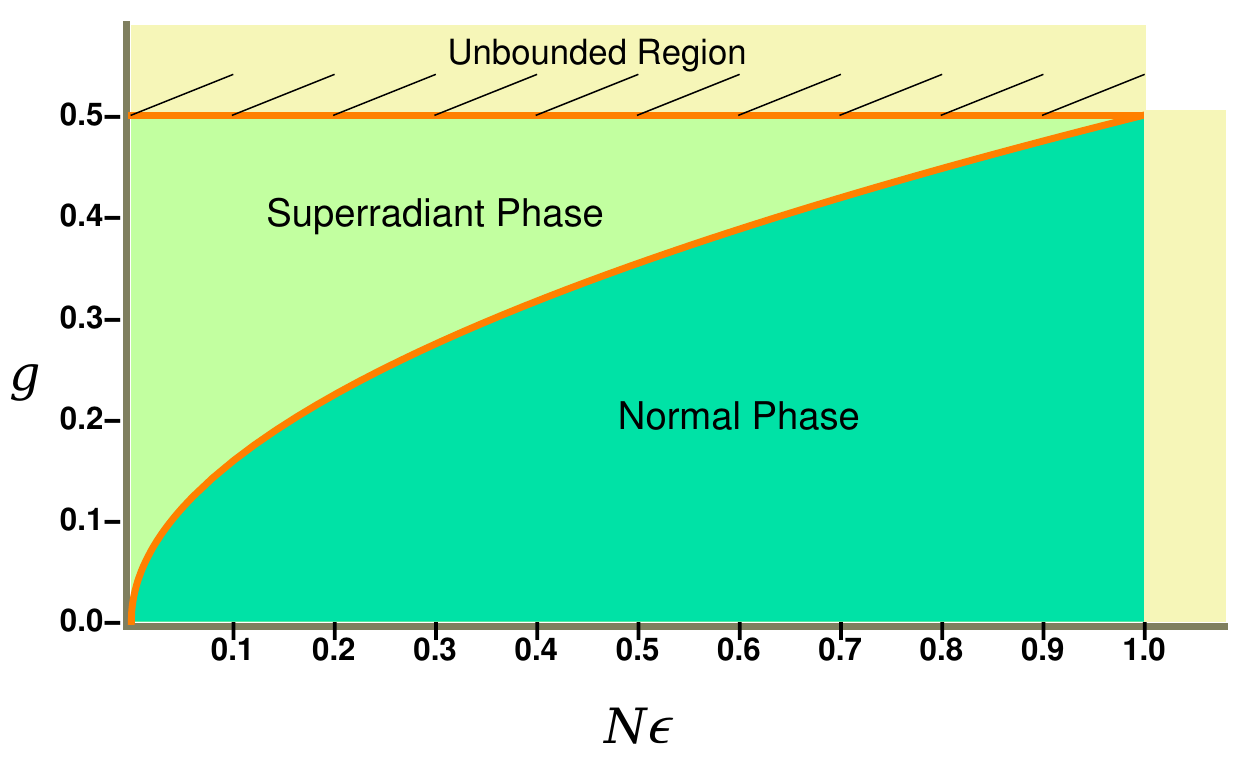}
\caption{\small{Phase diagram of the Two-Photon Dicke model in the mean-field approximation. $g$ and $N\epsilon$ are given in units of $\omega$. In the region where the coupling strength has a magnitude comparable to the fundamental frequency of the system, the model is no longer valid. This region is the unbounded region.}} 
\label{phasediag}
\end{figure}
The phase diagram of the Two-Photon Dicke Model shows two critical boundaries as opposed to one as in Dicke Model. The unbounded region beyond $g=\omega/2$ is the region where the energy levels collapse into a continuum \cite{PhysRevA.92.033817,PhysRevA.85.043805,Peng_2017}. The region in the superradiant phase very close to this boundary is of particular importance in our study.\\
Next, using $c_1$ and $c_2$, the frequency of elementary excitations of the bosons inside the cavity is,
\begin{equation}
\begin{split}
\omega_a & = 0 \text{ for  } g \leq g_t\\
 & =\pm \sqrt{\omega^2 - \Big[\frac{g\sqrt{N-|\beta_0|^2}}{N}(2\beta_0)\Big]^2} \text{ for  } g > g_t\\
\end{split}
\end{equation}
\section{Squeezing Dynamics in the Superradiant Phase}\label{sec:squeezing}

In both the phases, we assume the quadrature to be along a random direction. Hence, an arbitrary quadrature $\hat Q_\phi = \hat X \cos \phi + \hat P \sin \phi $ is defined, which is minimum for a certain value of $\phi$ (as per the formalism of Kitagawa and Ueda \cite{PhysRevA.47.5138}). The uncertainty in the quadrature $\hat Q_\phi$ is\\ $\Delta\hat Q_\phi = \sqrt{\left\langle \hat Q_\phi^2(t)\right\rangle - \left\langle \hat Q_\phi(t)\right\rangle^2}$. \\ The parameter $\zeta_Q(t) =\frac{ \Delta \hat{Q}_\phi(t)_{min}}{1/\sqrt{2}} $ quantifies quadrature squeezing as a function of time.
Now, $\hat Q_\phi$ can be written in terms of creation and annihilation operators as:
\begin{equation}\label{sq}
\begin{split}
\hat Q_\phi &= \hat X \cos \phi + \hat P \sin \phi \\
& = (\frac{\hat a + \hat a^{\dagger}}{\sqrt{2}})(\frac{e^{i \phi} +e^{- i \phi}}{2}) + i (\frac{\hat a^{\dagger}-\hat a }{\sqrt{2}})(\frac{e^{i \phi} -e^{- i \phi}}{2 i}) \\
\hat Q_ \phi & =  \sqrt{\frac{1}{2}}(\hat a^\dagger e^{i \phi} + \hat a e^{-i \phi})
\end{split}
\end{equation}
\begin{equation}\label{sqz}
\begin{split}
\zeta_Q(t)& =\sqrt{\left\langle \hat Q_\phi^2(t)\right\rangle-\left\langle \hat Q_\phi(t)\right\rangle^2}\\
 &= \sqrt{A_q(t)\cos 2 \phi + B_q(t) \sin 2 \phi + C_q(t) } \\
\end{split}
\end{equation}
Where,  
\begin{equation}
\begin{split}
A_q(t) & = \left\langle 0\middle | e^{i \hat H t}[\hat a^\dagger  \hat a^\dagger  + \hat a\hat a ] e^{-i \hat H t} \middle |0 \right\rangle \\
B_q(t) & =i\left\langle 0\middle | e^{i \hat H t}[ \hat a^\dagger  \hat a^\dagger  -  \hat a\hat a] e^{-i \hat H t} \middle |0 \right\rangle \\
C_q(t) & = \left\langle 0\middle | e^{i \hat H t}[2\hat a ^\dagger \hat a] e^{-i \hat H t} \middle |0 \right\rangle +1
\end{split}
\end{equation}
Now, the parameters $A_q(t)$,  $B_q(t)$ and  $C_q(t)$ are: 
\begin{equation}\label{12}
\begin{split}
A_q(t) &=\frac{2 \omega  g_{\beta } (\cos (2\omega_a t)-1)}{\omega_a ^2} \\
B_q(t) &= -\frac{2 g_{\beta } \sin (2\omega_a t)}{\omega_a } \\
C_q(t) &=\frac{\omega ^2-4 g_{\beta }^2 \cos (2\omega_a t)}{\omega_a ^2}
\end{split}
\end{equation}
 The order parameter $\beta$ in the normal phase is zero, hence the squeezing obtained is also zero \cite{PhysRevA.95.053854}.  Thus, we will only look at the variation of position and momentum quadratures with respect to time in the superradiant phase for a fixed beta by putting $\phi = 0$ and $\phi = \frac{\pi}{2}$ in \eqref{sqz} respectively.
\begin{figure}[H]
\centering
\includegraphics[width= 15cm,height=\textheight,keepaspectratio]{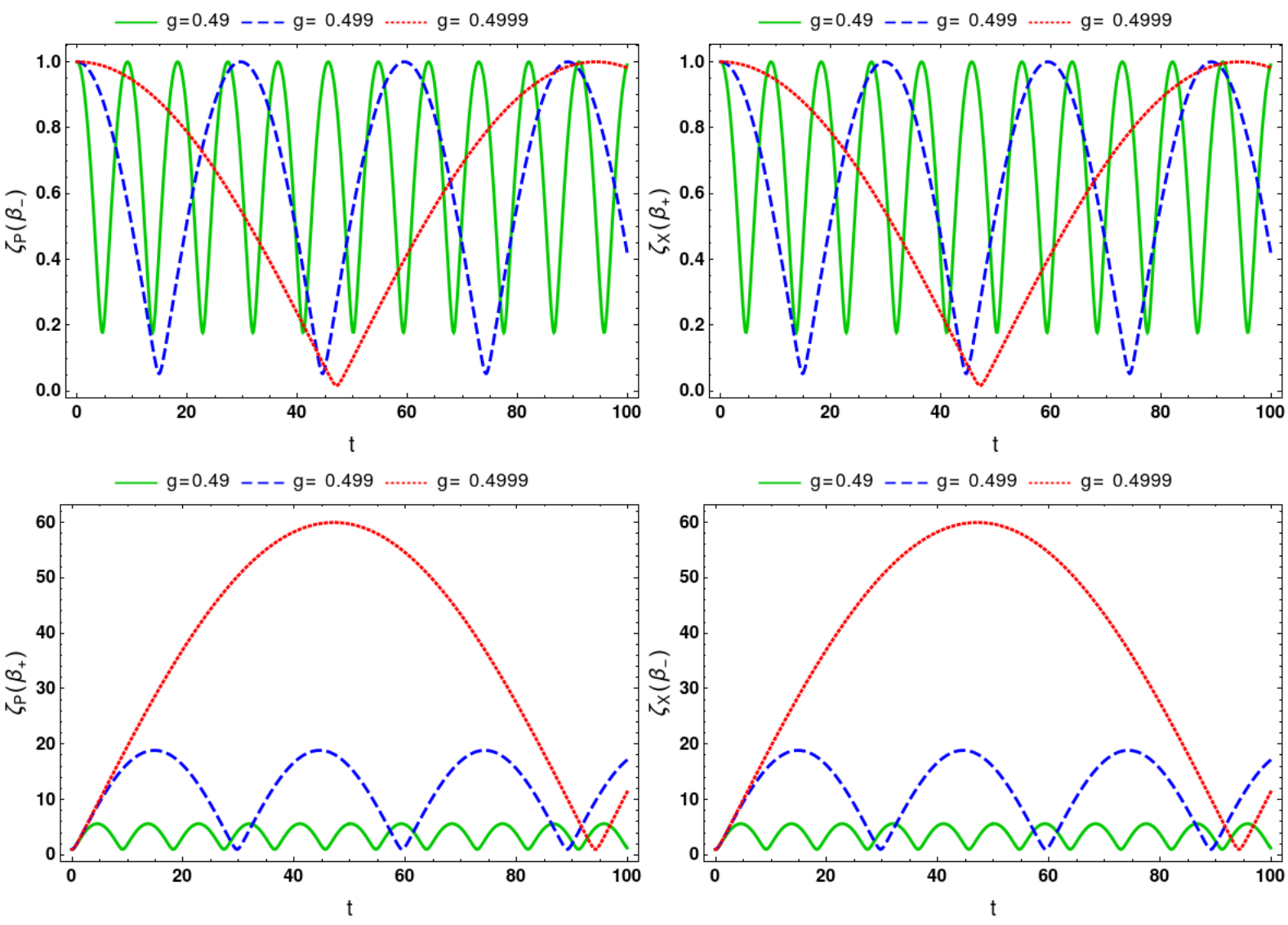}
\caption{\small{Squeezing parameter $\zeta_X $ (for Position quadrature) and $\zeta_P$ (for Momentum quadrature) for $\pm \beta_0$ in the superradiant phase of the Two-Photon Dicke model for $N = 1000,\omega = 1$ and $\epsilon=0.0008$.}}
\label{fig:2}
\end{figure}
In Figure \eqref{fig:2}, we see a complementary behavior, which is expected. That is, for a fixed beta, when a certain quadrature is squeezed, the other quadrature gets enhanced and vice versa. Also, quite clearly this squeezing is maximized as $g\to\omega/2$, which will be clearer in the next section.\\

\begin{figure}[H]
    \centering
    \subfloat[\centering]{{\includegraphics[width=7cm]{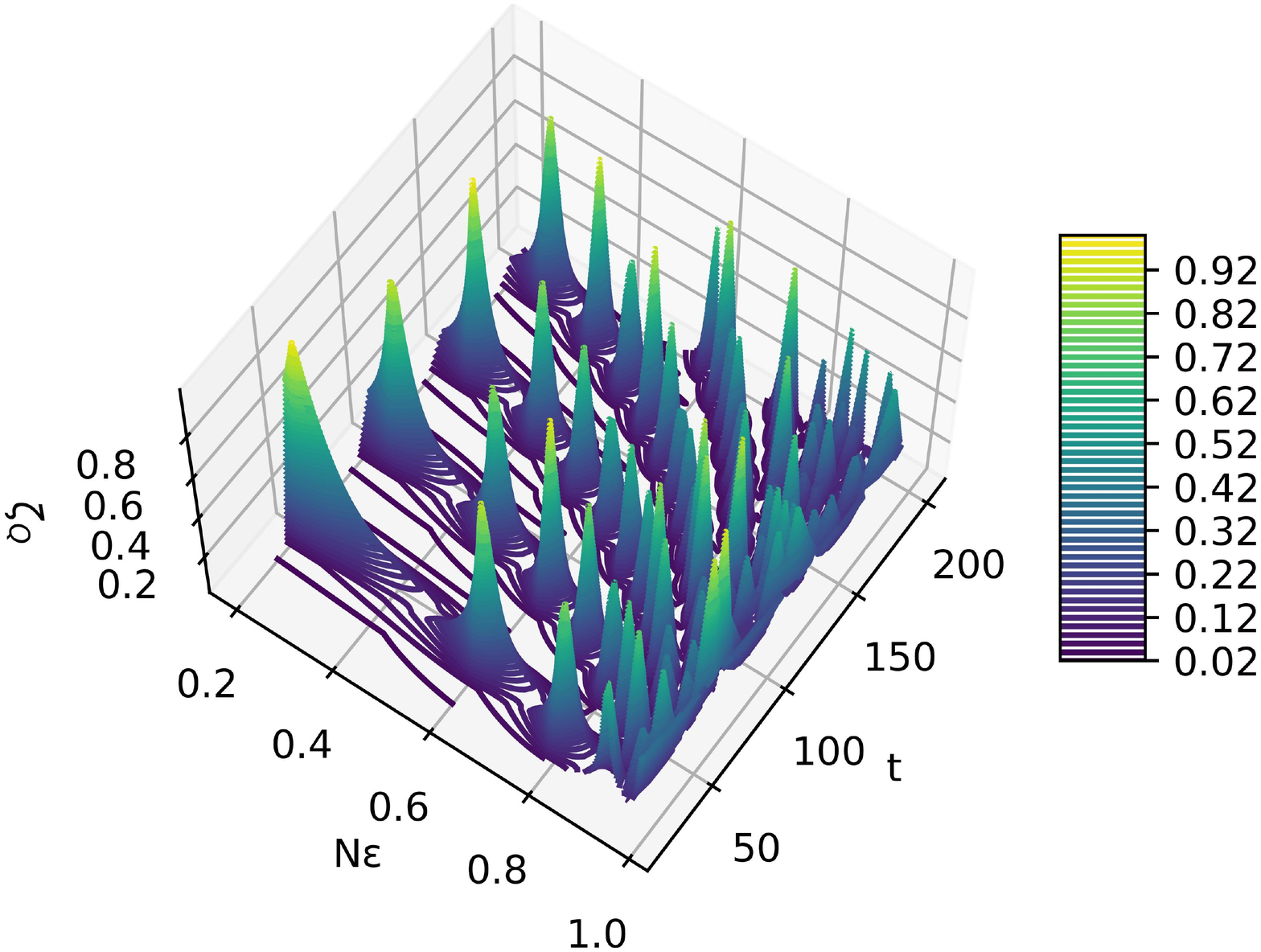} }}%
    \qquad
    \subfloat[\centering]{{\includegraphics[width=7cm]{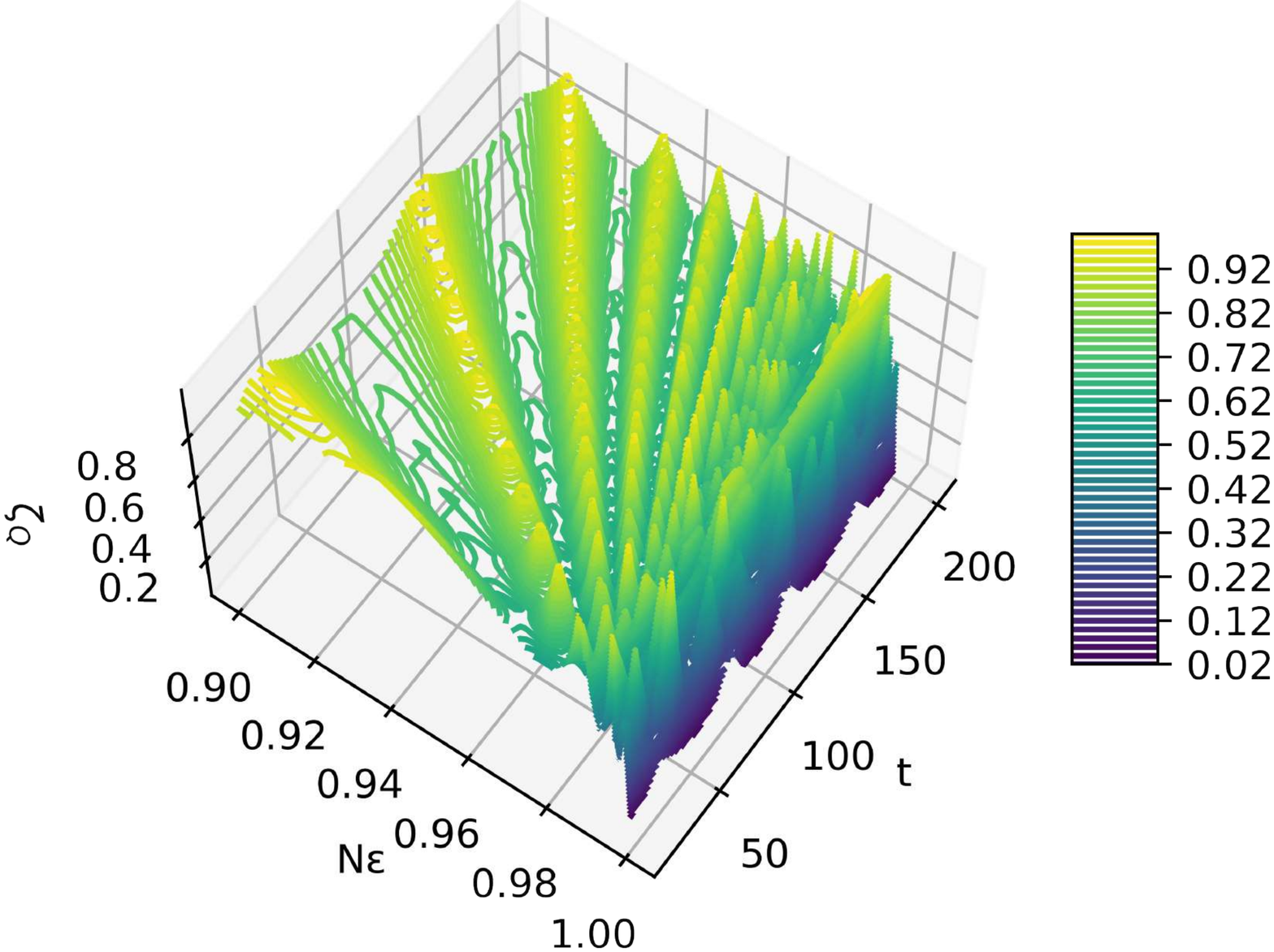} }}%
    \caption{\small{Generalised Quadrature Squeezing $\zeta_Q$ is plotted with respect to $N\epsilon$ for varying time where $t \in [40,200]$. Two cases we look at are (a) $g\to\omega/2$ where $N\epsilon\in[0.2,1]$ and (b) $g\to g_t$ where $N\epsilon\in[0.9,1]$,  in the superradiant phase of the Two Photon Dicke model. Other parameters are the same as in Figure \eqref{fig:2}.}}
\label{fig:5}
\end{figure}

Now we look at the generalised quadrature squeezing, first near the critical point ($g\to g_t$) and then farther away from $g_t$ and close to $\omega/2$ ($g\to\omega/2$). For this, in \eqref{sqz}, we consider  $\phi = \phi_{min} $ to obtain the minimum value of the squeezing parameter. Thus, $\Delta \hat Q_\phi(t) = \Delta \hat Q_\phi(t)_{min}$.
So, 
\begin{equation}\label{sq1}
\begin{split}
\zeta_Q(t)& = \sqrt{C_q(t) + e^{i\phi_{min}}\sqrt{A_q^2(t) + B_q^2(t)}} \\
& = \sqrt{C_q(t) -\sqrt{A_q^2(t) + B_q^2(t)}} 
\end{split}
\end{equation}
Where, $\phi_{min} = \frac{\pi}{2} + \frac{1}{2}\tan^{-1}\Big[\frac{B_q(t)}{A_q(t)}\Big]$. 
\begin{figure}[H]
    \centering
    \subfloat[\centering]{{\includegraphics[width=\textwidth]{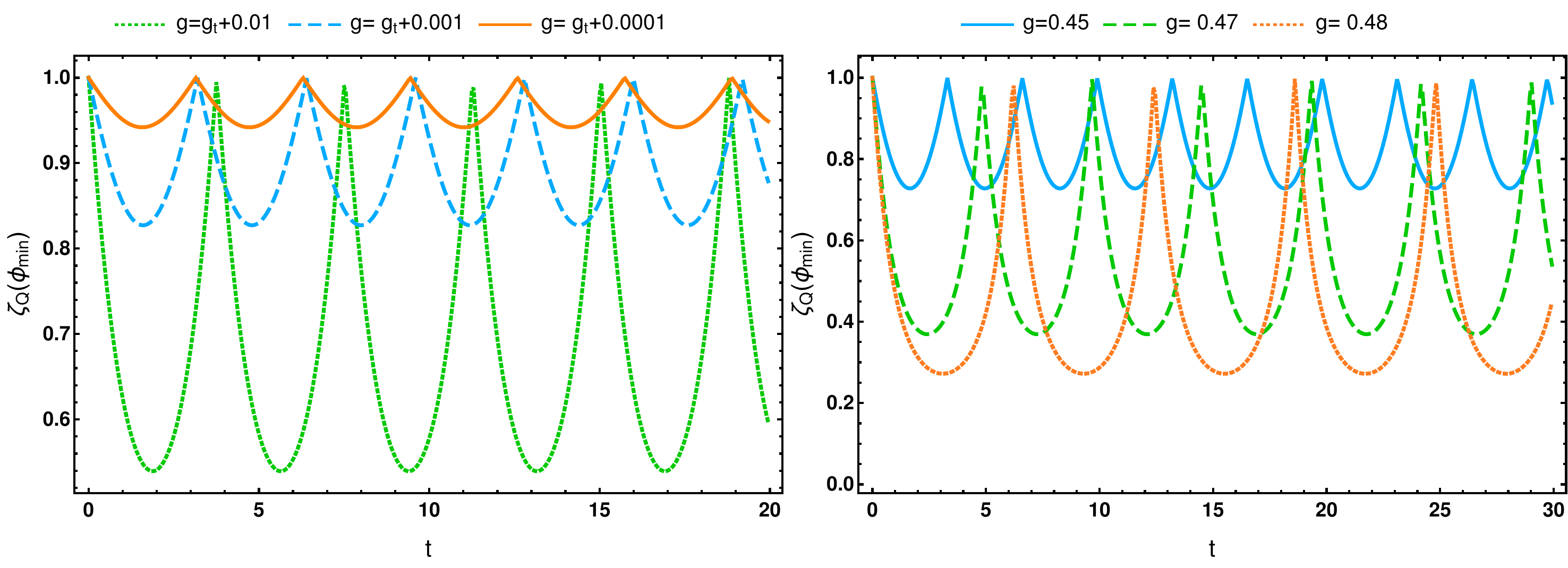} }}%
    \qquad
    \subfloat[\centering]{{\includegraphics[width=\textwidth]{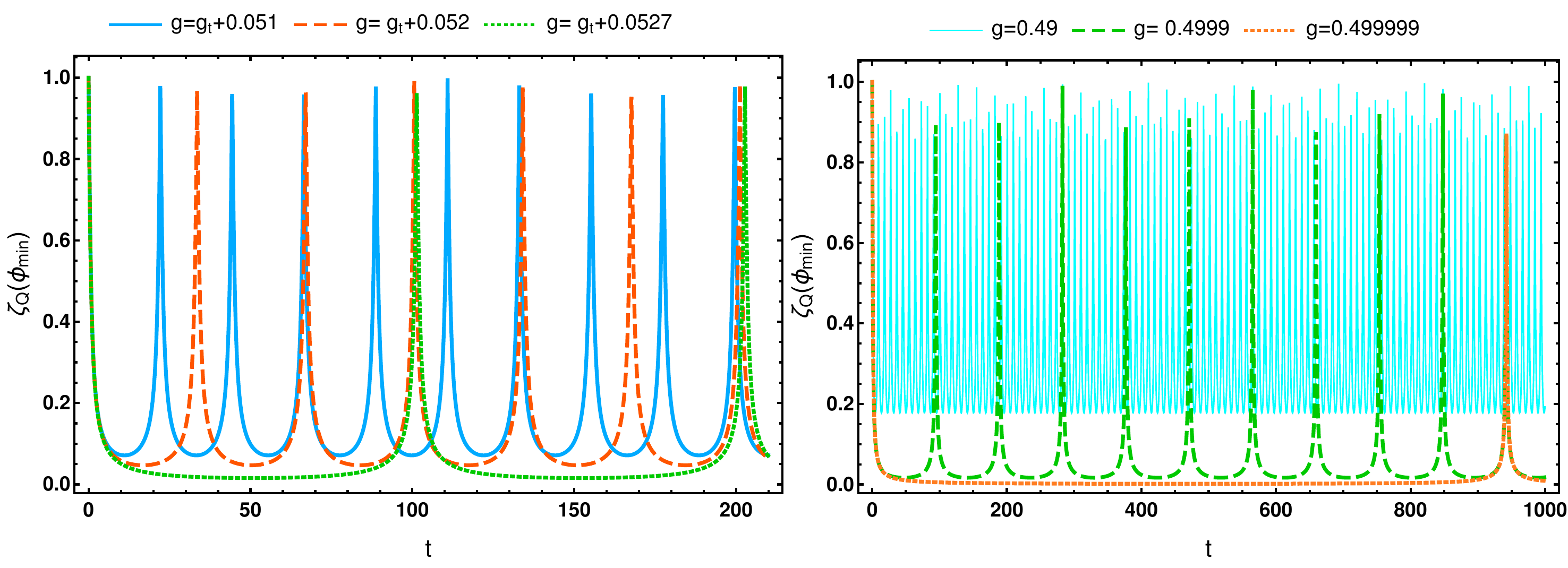} }}%
    \caption{\small{Generalised Quadrature Squeezing $\zeta_Q$ is plotted with respect to time. The two cases we look at are, (a) $g\to g_t$ and (b) $g\to\omega/2$ (taking $g_t=0.447$) in the superradiant phase of the Two-Photon Dicke model. 
Other parameters are the same as in Figure \eqref{fig:2}.}}
\label{fig:3}
\end{figure}
The degree of squeezing is quantified by $\zeta_Q^2(t)= \frac{ \Delta \hat{Q}_\phi(t)_{min}^2}{1/2} $. Squeezing of high magnitude is said to be achieved when the uncertainty in any one of the position or momentum quadratures is beyond the standard quantum limit (i.e. $\zeta_Q^2<1/2$). This corresponds to a magnitude significantly lower than $-10\log_{10}(1/2) \simeq 3$ dB \cite{MA201189}. Here, in Figure \eqref{fig:3}, we see that $\zeta_Q<<\frac{1}{\sqrt{2}}$ which means, squeezing is achievable well beyond the SQL (i.e 3 dB limit) for both cases (a) and (b).\\
The magnitude of squeezing obtained in (b)  $g\to\omega/2$ is much larger and persists for a longer time interval. For the case where $g=g_t+\Delta$, we will see in Figure \eqref{fig:5} that, if $\Delta\to0$, prolonged squeezing cannot be obtained even close to the unbounded region (i.e.  $g\to\omega/2$).  \\

Next, to find the optimum parameters to obtain maximized squeezing for a prolonged time interval, we vary $N\epsilon$ and hence $g_t$ \eqref{qpt} at any given time.

\begin{figure}[H]
\centering
\includegraphics[width= 10cm,height=\textheight,keepaspectratio]{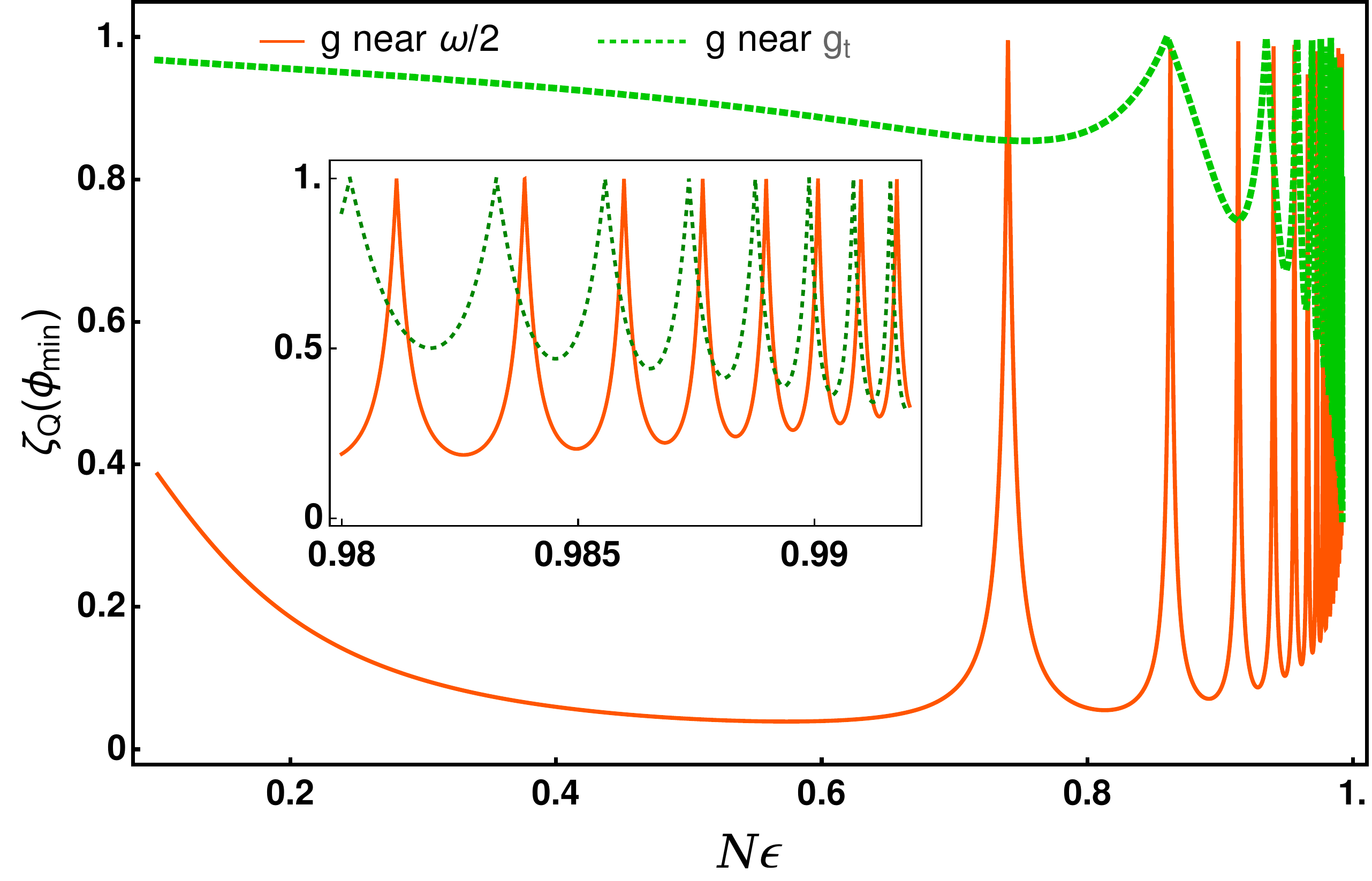}
\caption{\small{Generalised Quadrature Squeezing $\zeta_Q$ is plotted with respect to $N\epsilon$ at a fixed time $t = 100$. The orange line is for $g\to\omega/2$ and green dotted line for $g\to g_t$ in the superradiant phase of the Two Photon Dicke model. Other parameters are same as in Figure \eqref{fig:2}.}}
\label{fig:4}
\end{figure}
Here, we see that if $g$ is fixed at a value very close to $\omega/2$, we get a high magnitude of squeezing for lower values of  $N\epsilon$ and it shows an oscillatory behavior for higher values with minimum value of squeezing parameter gradually decreasing. On the other hand, very close to the critical point, the magnitude of squeezing obtained is very weak for lower values of $N\epsilon$ which increases significantly as $N\epsilon$ approaches 1. However, owing to its highly oscillatory behavior, an optimum value $\zeta_Q(\phi_{min})$ will be hard to maintain. For values of $N\epsilon$ very close to 1, the values of squeezing parameter converges as shown in the plot above. This is the point in the phase diagram shown in Figure \eqref{phasediag} where the phase boundary meets the $g=\omega/2$ line. \\
\hfill \newline
Let's now see how the squeezing parameter for these different $N\epsilon$ evolves with time. Then, we can finally conclude specifically whether a prolonged squeezing of high magnitude can be obtained close to the unbounded region or near the phase boundary.

We see above, how the squeezing parameter evolves with time for different values of $N\epsilon$. In (a), we fix $g$ such that $g\to\omega/2$ and in (b) we stay close to the quantum critical point which keeps varying with varying $N\epsilon$. Clearly, the magnitude of squeezing obtained is much higher for smaller values of $N\epsilon$ in (a) and larger values of $N\epsilon$ in (b), as shown in Figure \eqref{fig:4}. \\
Another striking feature is that, although squeezing is obtained in (b) at $N\epsilon\to1$, it is not as prolonged as in (a). This is because for $N\epsilon\to1$, $g_t\to \omega/2$. Hence, $g\simeq g_t$. As we showed in Figure \eqref{fig:3}, squeezing is prolonged only if $g\to \omega/2$ and $(g-g_t)\centernot \longrightarrow 0$. 

\subsection{Behavior of Squeezing near the Unbounded Region \label{sec:crit}}
To better understand the nature of the growth of the squeezing parameter $\zeta_Q$ near the unbounded region ($g \to \omega/2$), we plot the minimum value of the squeezing parameter for different values of $\delta$, where $g = \omega/2 - \delta$. We are only confined to the superradiant phase where squeezing is obtained in the Two-Photon Dicke Model.
\begin{figure}[H]
\centering
\includegraphics[width= 12.5cm,height=\textheight,keepaspectratio]{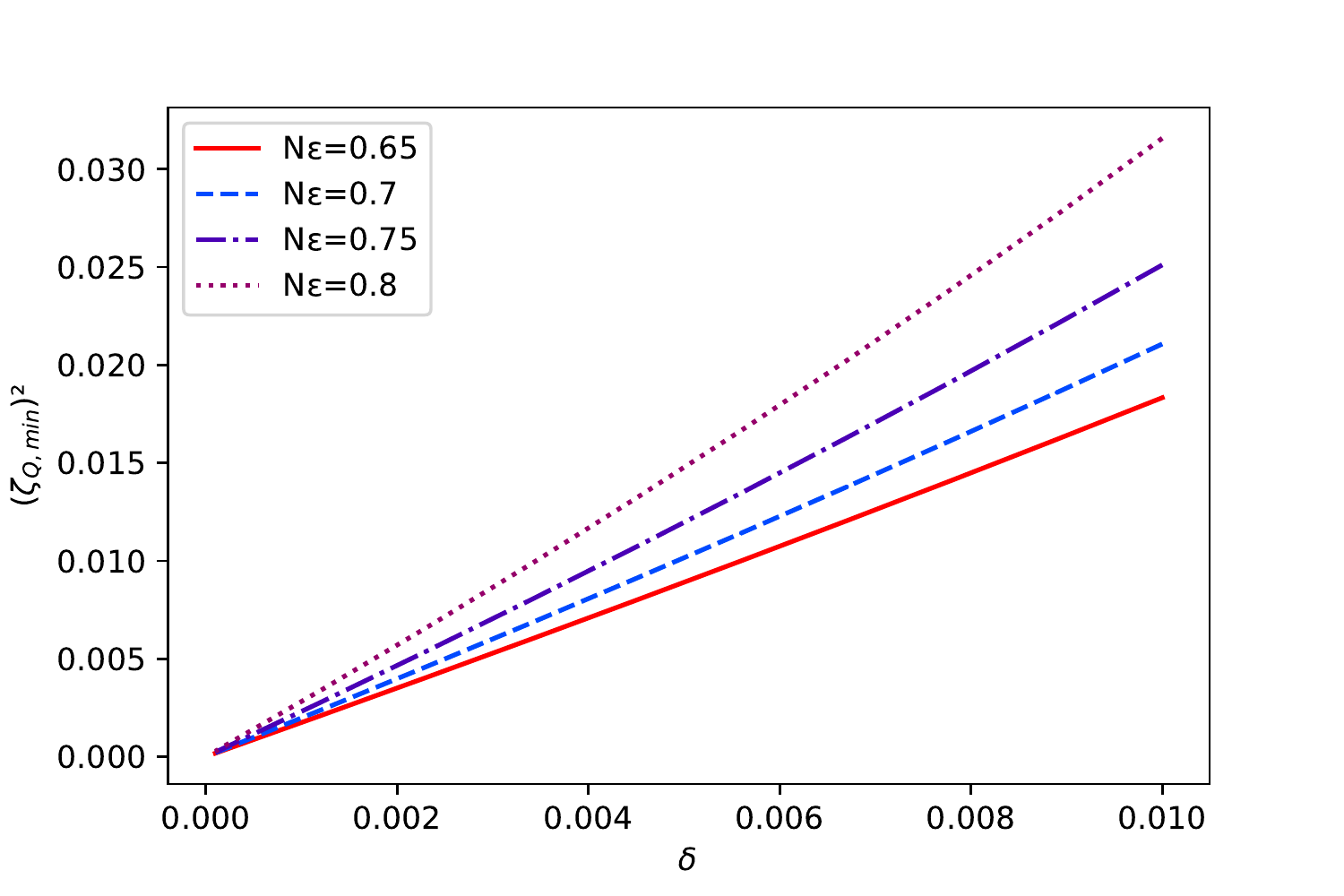}
\caption{\small{Critical behavior of the squeezing parameter in superradiant phase of the Two-Photon Dicke model. Other parameters are same as in Figure \eqref{fig:2}.}} 
\label{fig:6}
\end{figure}
Figure \eqref{fig:6} shows a linear relationship between the square of the minimum value of $\zeta_Q$ and $\delta$. This can be represented by the equation of a straight line passing through the origin,\\ $(\zeta_{Q,min})^2 = m \delta$, where m is the slope of the straight line. 
\begin{equation}\label{critical}
\begin{split}
(\zeta_{Q,min})^2 &= m \delta\\
\Rightarrow \zeta_{Q,min}&= \sqrt{m (\frac{\omega}{2}-g)}\\
\end{split}
\end{equation}
Thus, as in \eqref{critical}, the squeezing parameter behaves as $\zeta_{Q,min}\propto  \sqrt{\frac{\omega}{2}-g}$. The squeezing strength decreases with the increase in  the value of $N\epsilon$ as we move away from $g= \sqrt{\omega/2}$. This is in accordance with the behavior shown in  Figure \eqref{fig:4}. \\
Next, let us see how the squeezing parameter evolves with time for $g \to \omega/2$ (i.e. $\delta\to 0$). We can clearly see from \eqref{12}, the frequency of oscillation of the squeezing parameter is $2\omega_a $. Hence, the time period $T = \frac{\pi}{\omega_a}$. Now, expanding $\omega_a$ in a series and taking the leading order term, we get $\omega_a=2 \sqrt{\delta} \sqrt{\frac{\omega ^3}{ \omega ^2- N^2 \epsilon ^2}}$ i.e. $\omega_a \propto \sqrt{\delta}$. Thus, in the limit $\delta \to 0$, very close to the unbounded region the time period of oscillation of the squeezing parameter $\zeta_Q$ is proportional to $\delta^{-\frac{1}{2}}$.\\
Summarizing, the time period of oscillation of the squeezing parameter and its magnitude shows the following behavior in the limit  $\delta \to 0$,\\
\begin{equation}
\begin{split}
\text{Time Period of $\zeta_Q$} &\propto (\omega/2-g)^{-\frac{1}{2}}\\
\text{Magnitude of  $\zeta_Q$} &\propto (\omega/2-g)^{\frac{1}{2}}
\end{split}
\end{equation}
\newpage
\section{Conclusion}\label{sec:conclusion}
In summary, we discuss the nature and behavior of quadrature squeezing of photons trapped in a cavity near the quantum critical point and the unbounded region of the Two-Photon Dicke model. The model undergoes a quantum phase transition from the normal to the superradiant phase for a certain change in the order parameter $\beta$. However squeezing is obtained only in the Superradiant phase. The position and momentum quadratures of the photons, get squeezed for the positive and negative values of the order parameter $\beta=\beta_0$ respectively. We see that in the thermodynamic limit ($N\to \infty$), the Two-Photon Dicke Model shows enhanced squeezing near the quantum critical point as well as the unbounded region. However, the squeezing is prolonged provided $g$ is sufficiently larger than $g_t$ and close to $\omega/2$. The squeezing strength behaves as $(\omega/2-g)^{\frac{1}{2}}$ whereas the time of the squeezing grows as $(\omega/2-g)^{-\frac{1}{2}}$. To conclude, we have found that it is possible to achieve a high degree of squeezing for a prolonged time away from the quantum phase boundary and hence we try to find the optimum set of parameters to do so.

\bibliographystyle{ieeetr}
\bibliography{manuscript.bib}

\end{document}